\documentclass[9pt,twocolumn,twoside]{opticajnl}
\journal{opticajournal} % use for journal or Optica Open submissions

\setboolean{shortarticle}{true}
\usepackage{lineno}
\title{Cascaded multi-phonon stimulated Raman scattering near second-harmonic-generation in thin-film lithium niobate microdisk}

\author[1,$\dagger$]{Yuxuan He}
\author[1,2,$\dagger$]{Xiongshuo Yan}
\author[1]{Jiangwei Wu}
\author[1]{Xiangmin Liu}
\author[1,*]{Yuping Chen}
\author[1,3,4]{Xianfeng Chen}

\affil[1]{State Key Laboratory of Advanced Optical Communication Systems and Networks, School of Physics and Astronomy, Shanghai Jiao Tong University, Shanghai 200240, China
}
\affil[2]{Hangzhou Global Scientific and Technological Innovation Center, Zhejiang University, Hangzhou 310027, China}
\affil[3]{Jinan Institute of Quantum Technology, Jinan 250101, China}
\affil[4]{Collaborative Innovation Center of Light Manipulations and Applications, Shandong Normal University, Jinan 250358, China}

\affil[*]{corresponding author: ypchen@sjtu.edu.cn}

\begin{abstract}
High-quality microresonators can greatly enhance light-matter interactions and are excellent platforms for studying nonlinear optics. Wavelength conversion through nonlinear processes is the key to many applications of integrated optics. The stimulated Raman scattering process can extend the emission wavelength of a laser source to a wider range. Lithium niobate, as a Raman active crystalline material, has remarkable potential for wavelength conversion.
Here, we demonstrate the generation of cascaded multi-phonon Raman signals near the second-harmonic-generation peak in X-cut thin-film lithium niobate microdisk. Fine tuning of the specific cascaded Raman spectral lines has also been made by changing the pump wavelength. Raman lines can reach wavelength up to about 80 nm away from the SHG signal. We realize the SFG process associated with Raman signals in the visible range as well. Our work extends the use of WGM microresonators as effective optical upconversion wavelength converters in nonlinear optical applications.
\end{abstract}

\setboolean{displaycopyright}{false} % Do not include copyright or licensing information in submission.

\begin{document}
\setlength{\parskip}{0em}

\maketitle

% \section{Introduction}
High-Q factor microresonators are excellent platforms for the study of nonlinear optics by confining photons in the cavity for a long time through total internal reflection, which greatly enhances the light-matter interaction and increases the energy density of the optical field in the cavity\cite{lin2017nonlinear,zhu2021integrated,vahala2003optical}. Lithium niobate (LN) has been identified as a promising material with a high second-order susceptibility ($\chi^{(2)}$)\cite{lin2016phase}, high electro-optic (EO) coefficient\cite{wang2015high}, and a wide transparency window, making it an excellent contender for use in integrated photonics. The maturation of thin-film lithium niobate (TFLN) fabrication technologies over recent years is also paving the way for lithium niobate applications in next-generation photonic integrated circuits (PICs)\cite{luke2020wafer,he2019low}. The most advanced micro-nanofabrication techniques are now available, enabling the Q factor of TFLN microresonators to be increased to the order of $10^8$, approaching its theoretical limit\cite{gao2021broadband,wu2018lithium}.\par
The characteristics of TFLN microresonators, including high-Q factors and small volume, facilitate the realization of various strong nonlinear effects. Previously, effects such as second-harmonic generation (SHG)\cite{wang2017second,lu2019periodically}, third-harmonic generation (THG)\cite{liu2017cascading}, sum-frequency generation (SFG)\cite{strekalov2014optical} and spontaneous parametric down conversion (SPDC)\cite{chen2019efficient,xue2021effect} on TFLN microresonators have been extensively studied. Furthermore, phenomena such as optical parametric oscillation (OPO)\cite{lu2021ultralow,hwang2023mid}, optical frequency comb (OFC)\cite{gong2020near,yang20231550}, and cavity optomechanics\cite{cai2019acousto,hassanien2021efficient} have also been extensively demonstrated on TFLN microresonators. TFLN microresonators are also becoming increasingly important in optical amplification\cite{wu2024efficient}, optical communications\cite{xu2020high} and optical sensing\cite{wang2024enhanced,wang2024lpr}.\par
Stimulated Raman scattering (SRS), a third-order nonlinear nonparametric process, has been demonstrated on a variety of photonic integration platforms, including silicon\cite{yang2005erbium}, silicon dioxide\cite{choi2019low}, diamond\cite{latawiec2015chip}, and aluminum nitride\cite{liu2017integrated}. By aligning the pump wavelength and the Stokes light wavelength, which is related to the phonon vibrational frequency, with the cavity modes, SRS can provide new wavelengths that are different from the pump. This contributes to the realization of integrated Raman lasers with low pump levels in the continuous-wave (cw) regime. However, as a Raman-active crystalline material with different polarization configurations and multiple strong vibrational phonon branches\cite{gorelik2019raman,johnston1968stimulated}, the Raman effect in LN integrated photonic devices has not been extensively studied. Recent works have investigated cascaded Raman lasing\cite{yu2020raman,zhao2023integrated} in TFLN microresonators and the effect of SRS on the Kerr comb formation\cite{yu2020raman,zhao2023widely}. However, the majority of the works is concentrated on the SRS in the vicinity of the optical communication band (C-Band). In order to ascertain the potential of the LN Raman effect for wavelength conversion, it is necessary to investigate the SRS in other bands of the LN microresonators.\par
In this paper, we demonstrate the generation of cascaded multi-phonon Raman signals near SHG peak and related cascaded SFG processes by modal-phase-matching condition in an X-cut high-Q factor LN microdisk cavity under cw optical pumping at around 1543 nm. We fabricated TFLN microdisk cavity with Q factors higher than $8\times10^5$. The high Q and small mode volume of the WGM modes in the microdisk compensate for the small spatial mode overlap between the interacting modes, allowing the SRS effect and cascaded nonlinear effects to be observed. The spectrum of multi-phonon Raman signals and their cascaded SFG signals can be changed by tuning the pump frequency within a small interval. Furthermore, we observed the generation of multi-color visible light in the LNOI microdisk cavity under the optical microscope.

\begin{figure}[t]
\centering\includegraphics[width=\linewidth]{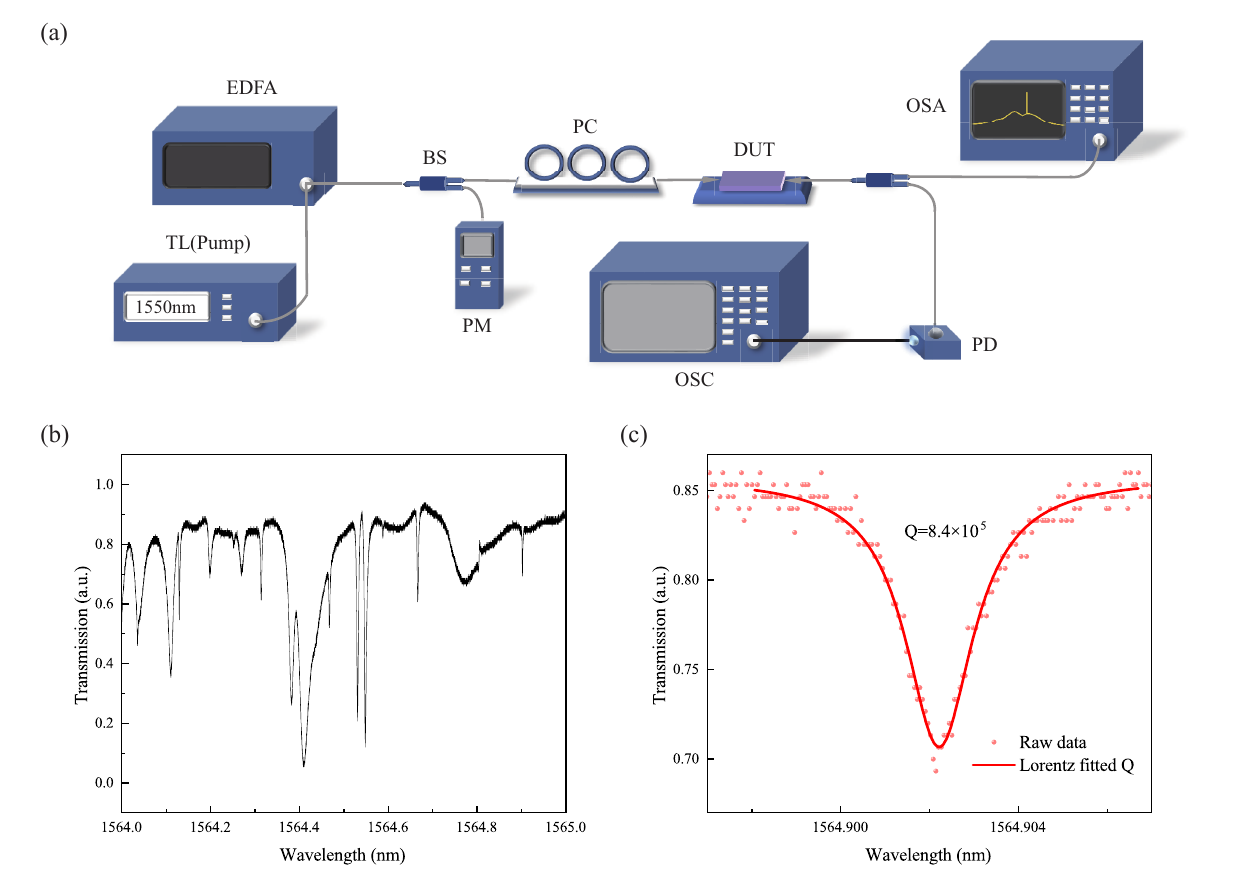}
\caption{~~~(a) Schematic illustration of the experimental setup. (b) Normalized transmission spectrum of the LNOI microdisk at telecommunication wavelengths. (c) The Lorentz fitting (red curve) reveals a Q factor of $8.4\times10^5$. EDFA, erbium-doped fiber amplifier; BS, beam splitter; PC, polarization controller; PD, photodetector; OSA, optical spectrum analyzer; OSC, oscilloscope.}
\end{figure}

% \section{RESULTS}

We fabricate the LN microdisk resonator with a radius of 100 $\mu m$ using X-cut TFLN with a LN layer of 300 nm on the top of a silica layer with a thickness of 2 $\mu m$, and the bottom layer is a LN substrate with a thickness of 500 $\mu m$. We use photolithography-assisted chemo-mechanical polishing (CMP) to obtain smoother sidewalls for the microdisk and thus higher Q factors. First, a layer of chromium (Cr) is deposited on the LN film by evaporation, and photoresist is spin-coated on the surface of the film, followed by exposure of a circular pattern using ultraviolet (UV) lithography. The microdisk with a flat surface and smooth sidewalls is obtained through a process of two wet etching and CMP. Finally, the silica under the LN microdisk was etched with the buffered oxide etch (BOE) method to form pillars to support the suspended microdisk. 

\begin{figure}[t]
\centering\includegraphics[width=\linewidth]{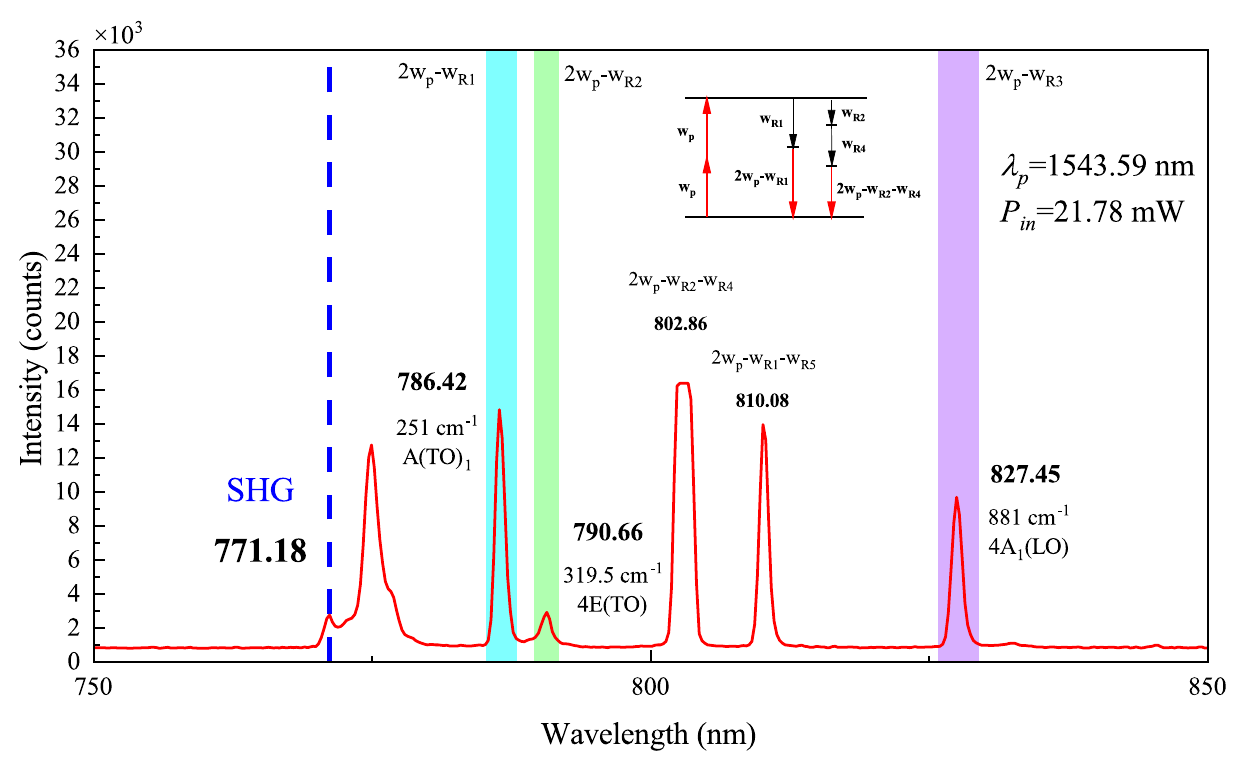}
\caption{~~~Experimental results of SRS. Spectrum of the generated SRS near SHG at pump wavelength of 1543.59 nm. The input power is fixed at 21.78 mW. The peak of SHG is at 771.18 nm. By comparing the wavelength of these spectral lines to the SHG peak, these lines are recognized as Raman and cascaded Raman lines. The first-order Raman signals at 786.42 nm, 790.66 nm and 827.45 nm decorated with blue, green and purple pillars are associated with Raman active phonons of ($1A_1 ~TO$), ($4E ~TO$), and ($4A_1 ~LO$), respectively. The spectral lines at 802.86 nm and 810.08 nm are cascaded Raman signals associated with the first-order Raman signals mentioned above. The energy level diagram in the figure explains the cases of frequency conversion in our nonlinear processes.}
\end{figure}

\begin{table}[ht]
\centering
\caption{\bf The Relationship between Raman Spectral Lines and Raman Active Phonons in Fig. 2}
\begin{tabular}{ccc}
\hline
$\lambda~[\mu m]$ & $\tilde{\upsilon} = \frac{1}{\lambda_{SHG}}-\frac{1}{\lambda}~[cm^{-1}]$ & Raman Active Phonons \\
\hline
786.42 & 251.3 & $(1A_1 ~TO)$ \\
790.66 & 319.5 & $(4A ~TO)$ \\
827.45 & 881.8 & $(4A_1 ~LO)$ \\
802.86 & 511.7 & $(4A ~TO)~and~(1E ~LO)$ \\
810.08 & 622.7 & $(1A_1 ~TO)~and~(5E ~TO)$ \\
\hline
\end{tabular}
  \label{tab:shape-functions}
\end{table}

Our experimental setup is schematically depicted in Fig. 1(a). The transmission spectrum characterizing the optical modes of the microdisk was measured before our experiment. To avoid thermal and nonlinear optical effects, the transmission spectrum is scanned using a very low input power over the wavelength range from 1520 nm to 1570 nm, as shown in Fig. 1(b).
The high Q-factor of the mode around 1564.90 nm is estimated to be $8.4\times10^5$ by the Lorentz fitting (red solid line).
Due to the lack of a tunable laser source, it is not possible to accurately measure the Q-factor in the visible range. However, we can expect the Q-factor to be similar to that of the C-band.

We first investigate the SHG process in our microdisk. We perform a scanning operation on the pump with the wavelength range from 1520 nm to 1570 nm. Within the scanning range, high intensity SHG signals can be generated at multiple pump wavelengths, indicating that the phase-matching condition of SHG is widely satisfied in the microdisk.
It is due to the fact that our microdisk has a large radius, which leads to the emergence of a large number of higher-order modes. It facilitates the realization of modal-phase-matching condition for the SHG process.

\begin{figure}[t]
\centering\includegraphics[width=\linewidth]{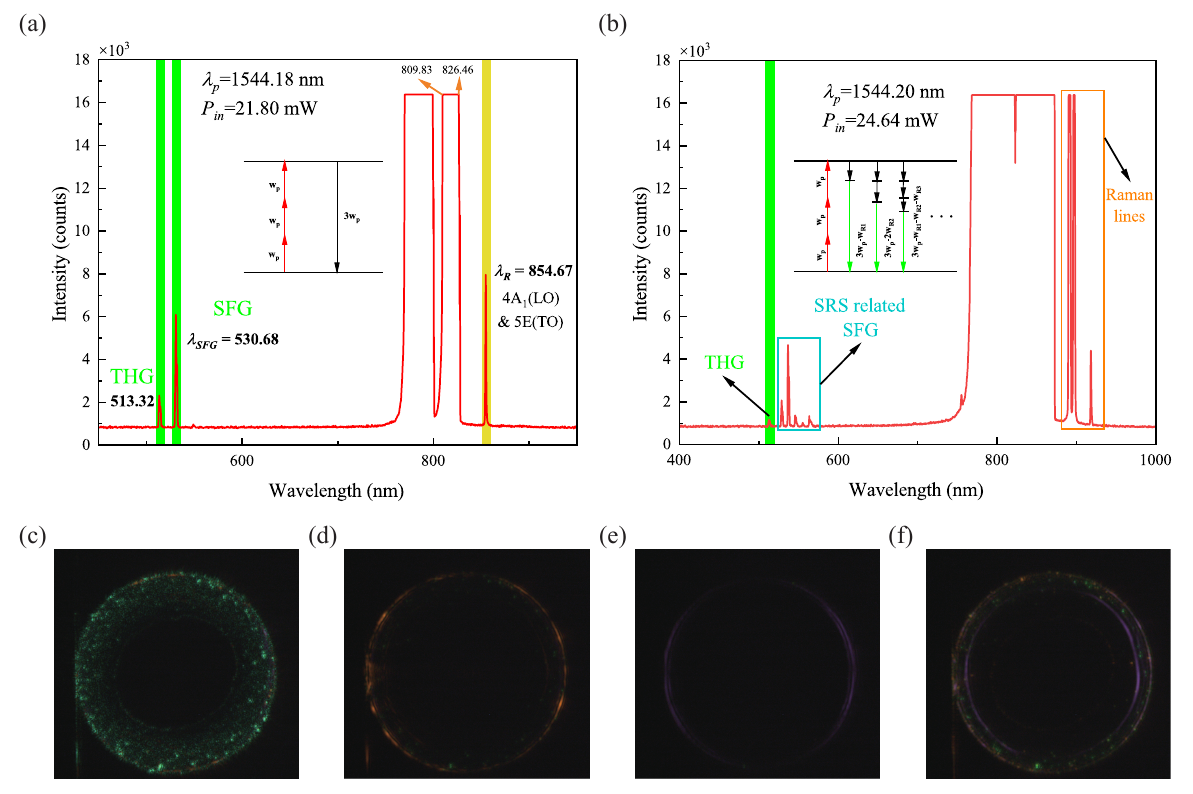}
\caption{~~~Raman lines emission accompanied by SFG process. (a) Two spectral lines appear in the visible range which are decorated with green pillars. The THG line is denoted as the pillar on the left while the SFG line is denoted as the pillar on the right. (b) Spectrum of the generated SFG process while slightly changing the pump wavelength and power. The increase in the number of spectral lines in the visible range represents the enhancement of the SFG process. The energy level diagram in the figure explains the cases of frequency conversion in THG and SFG processes. These spectral lines reveal the multiple parametric processes that are taking place in our experiment. Optical micrographs of (c) green light emission from the microdisk captured by optical microscope, (d) the red light emission, (e) the purple light emission, and (f) the mixed emission of these different color.}
\end{figure}

Next, we tune the pump wavelength to 1543.59 nm and fix the power at 21.78 mW, and observe multi-phonon Raman spectral lines near the SHG peak as shown in Fig. 2. The SHG peak appears at 771.18 nm. By comparing the wavelength of these spectral lines to the wavelength of the SHG peak and calculating the difference value of the wavenumber $\tilde{\upsilon} = \frac{1}{\lambda_{SHG}}-\frac{1}{\lambda}$, the frequency shifts of the spectral lines at 786.42 nm, 790.66 nm, and 827.45 nm are 251 $cm^{-1}$, 319.5 $cm^{-1}$ and 881 $cm^{-1}$. Referring to the Raman spectra of the LN crystal\cite{gorelik2019raman}, we can find that the spectral lines at 786.42 nm, 790.66 nm, and 827.45 nm are the first-order Raman signals near SHG correlated to the Raman-active phonons of ($1A_1 ~TO$), ($4E ~TO$), and ($4A_1 ~LO$). Here A and E are the two symmetric polarization directions of the LN crystal, and TO and LO are the transverse and longitudinal optical phonon modes of the LN crystal. These three peaks are labeled by blue, green and purple pillars in Fig. 2, respectively. In addition, we also observe spectral lines which appear at 802.86 nm and 810.08 nm. By comparing the wavelength of the spectral line at 802.86 nm to the first-order Raman signal at 790.66 nm, the frequency shift is 192 $cm^{-1}$ which is corresponded to Raman active phonon of ($1E ~LO$). Similarly, by comparing the wavelength of the spectral line at 810.08 nm to the  first-order Raman signal at 786.42 nm, the frequency shift is 371 $cm^{-1}$ which is corresponded to Raman active phonon of ($5E ~TO$). The spectral lines at 802.86 nm and 810.08 nm can be seen as cascaded Raman signals associated with the first-order Raman signals mentioned above. The relationship between Raman spectral lines and Raman active phonons are shown in Table 1.
It is unlikely that these first-order and cascaded Raman signals are directly triggered by the SHG signal due to the low intensity of the SHG signal. Therefore, we believe that the multi-color Raman signals observed in our experiment originates from the Raman signals generated by the pump light in the C-Band through SHG process. Due to the limitations of the range of our spectrometer, the Raman signals near the pump wavelength cannot be observed directly. In fact, we also observe signal peaks at 845.53 nm and 893.05 nm. They could be generated by the same reasons mentioned above. However,  we do not introduce them here due to the low intensity measured.

\begin{figure}[t]
\centering\includegraphics[width=\linewidth]{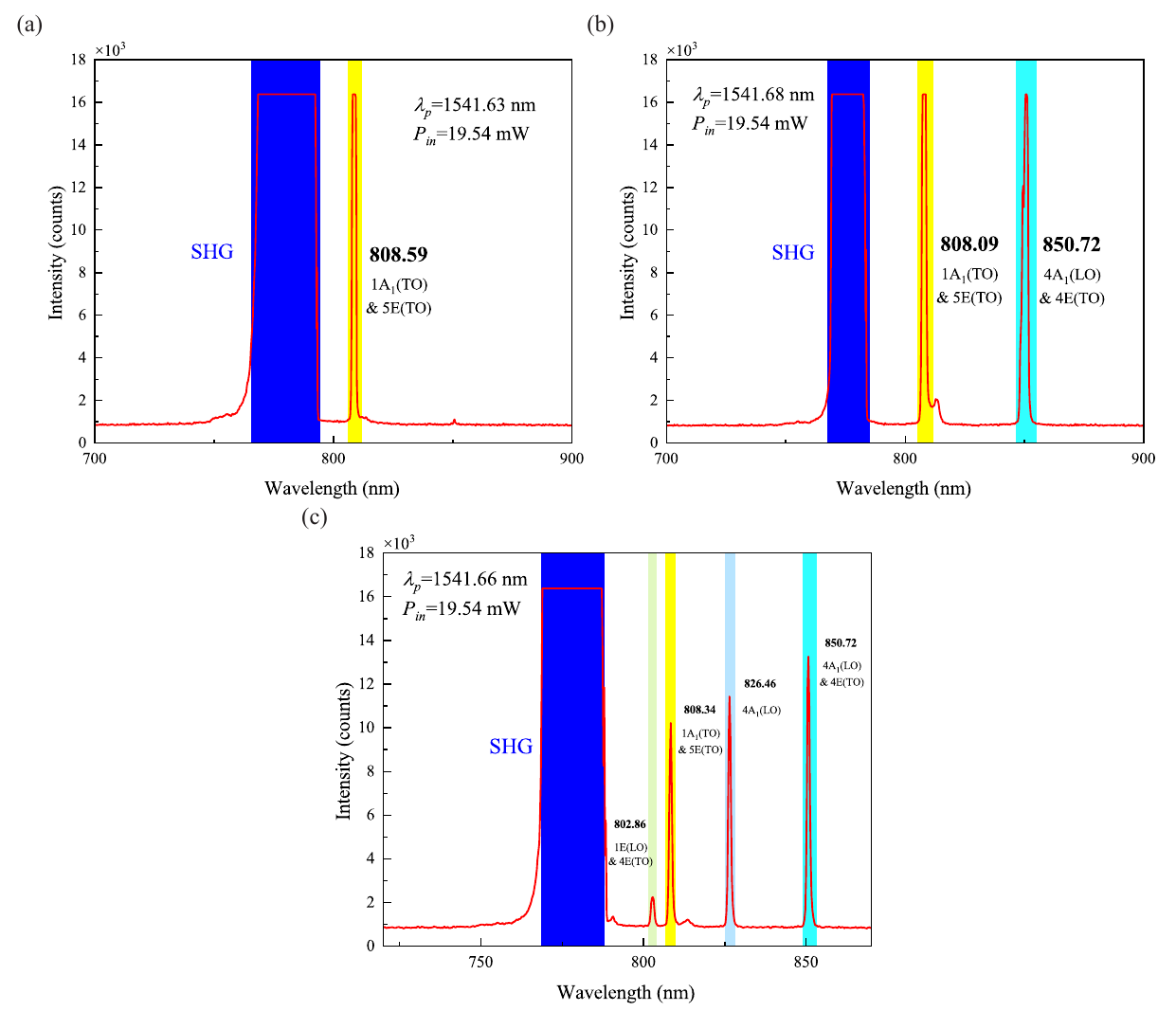}
\caption{~~~Fine tuning of the Raman peaks and cascaded Raman peaks. The input power is fixed at 19.54 mW. (a) Spectrum of the generated Raman spectral line and SHG spectral line at pump wavelength of 1541.63 which shows a "flat top" SHG signal and a Raman spectral line at 808.59 nm. The SHG signal is decorated with blue pillar. (b) and (c) Spectrum of the generated Raman peaks and cascaded Raman peaks when the pump is tuned to 1541.68 nm and 1541.66 nm. These Raman signals are decorated with shallow color pillars.}
\end{figure}

While keeping the power unchanged and further tuning the pump wavelength to 1544.18 nm, we observe Raman lines emission near SHG signal accompanied by THG and SFG processes as shown in Fig 3(a). The “flat top” spectral line on the left represents the high intensity of SHG, and the “flat top” spectral line on the right probably represents the high intensity of Raman line. It shows that the intensity of the spectral lines exceeds the detection range of our spectrometer. The spectral line at 854.67 nm may be the cascaded Raman signal which is related to the Raman active phonons of ($4A_1 ~LO$) and ($5E ~TO$) according to our analysis. More importantly, we record the SFG and THG signals in the visible range. They are decorated with green pillars as shown in Fig. 3. The spectral line located at 530.68 nm is generated by the SFG process of the pump light and the Raman signal near SHG which means the satisfaction of the modal-phase-matching condition.
Further, we record the spectrum which is shown in Fig. 3(b) after tuning the pump wavelength to 1544.20 nm and increasing the pump power to 24.64 mW. Not only does the intensity of the SHG and Raman spectral lines near 800 nm continue to increase, but spectral lines also appear near 900 nm. These spectral lines may be generated by the SFG processes between the Raman spectral lines produced directly by the pump in the C-band. It is worth noting that there is a significant increase in the number of SFG spectral lines in the visible range. This indicates that the modal-phase-matching condition between the fundamental modes near the pump and the Raman signals near SHG gets further satisfaction under the adjustment of the pump wavelength and power. But it is hard to figure out which Raman line is involved in the process due to the "flat top" phenomenon. 
We also observe the spatial mode intensity profile recorded by the optical microscope. The emission of green light and red light is demonstrated in Fig. 3(c) and Fig. 3(d), respectively. It should be noted that purple light also appears under the optical microscope but we cannot record the corresponding frequency components by our spectrometer. This is mainly due to the fact that we use only one tapered fiber for input and output. The signals at the shorter wavelength cannot be efficiently coupled out from the microdisk cavity through the tapered fiber designed for telecom band coupling. Fig. 3(f) demonstrates the mixed emission of the different color we mentioned above which shows the multi-color spectrum we obtained from our experiment.

Finally, we tune the pump wavelength to 1541.63 nm and the pump power to 19.54 mW, and the spectrum is shown in Figure. 4(a). The “flat top” spectral line decorated with dark blue pillar represents the high intensity of SHG. We exhibit only one Raman spectral line at 808.59 nm which may be a cascaded signal related to Raman active phonons of ($1A_1 ~TO$) and ($5E ~TO$). Forward tuning the pump wavelength to 1541.68 nm, we record the spectrum shown in Fig. 4(b). Besides the Raman line at 808 nm, we obtain another cascaded Raman line at 850.72 nm which may be related to Raman active phonons of ($4A_1 ~LO$) and ($4E ~TO$).
Reverse tuning the pump wavelength to 1541.66 nm, we record the spectrum shown in Fig. 4(c). Besides the Raman lines mentioned above, we obtain the third Raman line at 826.46 nm which may be related to Raman active phonons of ($4A_1 ~LO$) and the forth cascaded Raman line at 802.86 nm which may be related to Raman active phonons of ($4E ~TO$) and $1E ~LO$). As shown in Fig. 4, by changing the pump wavelength, we can make fine tuning of the specific Raman and cascaded Raman spectral lines in the microdisk. This is essentially due to the adjustment of the modal-phase-matching condition.
It is worth mentioning that the spectral width of the "flat top" spectral line of SHG may cause the Raman spectral lines under 800 nm to be masked, such as the Raman lines.

Our experiment has the characteristic of broadband matching and tunable operation. This is mainly attributed to the unique quasi-phase-matching condition in the X-cut microdisk and the multi-resonance condition of the microdisk.
As transverse-electrically polarized light waves in X-cut microdisk undergo rotational crystal orientation during propagation, the effective refractive index and effective nonlinear coefficient oscillate with the azimuthal angle of the microdisk\cite{zhao2023integrated, lin2019broadband}. This results in a relaxation of the stringent conditions for quasi-phase-matching. At the same time, the complex higher-order modes in the microdisk form a multiple-resonance background.
All these conditions make the phase-matching conditions easy to fulfill and adjust for the stimulated Raman scattering in our experiments. However, the spatial modal overlap factor between these interacting modes may be small\cite{zhao2023integrated}, which can be compensated by the high quality factor of the microdisk.

In conclusion, in this letter, we observed the on-chip multi-phonon cascaded Raman signals near the SHG wavelength range. Cascaded Raman signals and related SFG signals are generated in a high Q factor X-cut TFLN microdisk cavity by satisfying modal-phase-matching condition with a cw pump operating in the C-band at an input power of about 20 mW. This work expands the way for the realization of on-chip wavelength conversion over an ultra-wide wavelength range.

\begin{backmatter}
\bmsection{Funding} This work was supported by the National Natural Science Foundation of China (Grant Nos. 12134009), and Shanghai Jiao Tong University (SJTU) (Grant No. 21X010200828).

\bmsection{Acknowledgments} The authors thank the Center for Advanced Electronic Materials and Devices (AEMD) of Shanghai Jiao Tong University (SJTU) for the supports in device fabrications.

\bmsection{Disclosures} The authors declare no conflicts of interest.

\bmsection{Data availability} Data underlying the results presented in this paper are not publicly available at this time but may be obtained from the authors upon reasonable request.

\end{backmatter}

% \section{References}

% Bibliography
\bibliography{ref}

% Full bibliography added automatically for Optics Letters submissions; the following line will simply be ignored if submitting to other journals.
% Note that this extra page will not count against page length
% \bibliographyfullrefs{ref}

%Manual citation list
%\begin{thebibliography}{1}
%\bibitem{Zhang:14}
%Y.~Zhang, S.~Qiao, L.~Sun, Q.~W. Shi, W.~Huang, %L.~Li, and Z.~Yang,
 % \enquote{Photoinduced active terahertz metamaterials with nanostructured
  %vanadium dioxide film deposited by sol-gel method,} Opt. Express \textbf{22},
  %11070--11078 (2014).
%\end{thebibliography}

% Please include bios and photos of all authors for aop articles
\ifthenelse{\equal{\journalref}{aop}}{%
\section*{Author Biographies}
\begingroup
\setlength\intextsep{0pt}
\begin{minipage}[t][6.3cm][t]{1.0\textwidth} % Adjust height [6.3cm] as required for separation of bio photos.
  \begin{wrapfigure}{L}{0.25\textwidth}
    \includegraphics[width=0.25\textwidth]{john_smith.eps}
  \end{wrapfigure}
  \noindent
  {\bfseries John Smith} received his BSc (Mathematics) in 2000 from The University of Maryland. His research interests include lasers and optics.
\end{minipage}
\begin{minipage}{1.0\textwidth}
  \begin{wrapfigure}{L}{0.25\textwidth}
    \includegraphics[width=0.25\textwidth]{alice_smith.eps}
  \end{wrapfigure}
  \noindent
  {\bfseries Alice Smith} also received her BSc (Mathematics) in 2000 from The University of Maryland. Her research interests also include lasers and optics.
\end{minipage}
\endgroup
}{}

\end{document}